\documentclass[twocolumn]{aastex6}
\usepackage{threeparttable,rotating,amsmath}
\usepackage{amssymb}
\usepackage{url}
\usepackage{natbib}
\usepackage{amsmath}
\usepackage{url}
\usepackage{graphicx}
\usepackage[]{units}
\usepackage{url}
\usepackage{color}

\def\b2{B2~1215+30}
\def\lat{\textit{Fermi}-LAT}

\begin{document}

\title{A luminous and isolated gamma-ray flare from the blazar \b2}
\shorttitle{A luminous TeV flare from \b2}

\author{
A.~U.~Abeysekara\altaffilmark{1},
S.~Archambault\altaffilmark{2},
A.~Archer\altaffilmark{3},
W.~Benbow\altaffilmark{4},
R.~Bird\altaffilmark{5},
M.~Buchovecky\altaffilmark{6},
J.~H.~Buckley\altaffilmark{3},
V.~Bugaev\altaffilmark{3},
K.~Byrum\altaffilmark{7},
M.~Cerruti\altaffilmark{4},
X.~Chen\altaffilmark{8,9},
L.~Ciupik\altaffilmark{10},
W.~Cui\altaffilmark{11,12},
H.~J.~Dickinson\altaffilmark{13},
J.~D.~Eisch\altaffilmark{13},
M.~Errando\altaffilmark{3,14},
A.~Falcone\altaffilmark{15},
Q.~Feng\altaffilmark{11},
J.~P.~Finley\altaffilmark{11},
H.~Fleischhack\altaffilmark{9},
L.~Fortson\altaffilmark{16},
A.~Furniss\altaffilmark{17},
G.~H.~Gillanders\altaffilmark{18},
S.~Griffin\altaffilmark{2},
J.~Grube\altaffilmark{10},
M.~H\"{u}tten\altaffilmark{9},
N.~H{\aa}kansson\altaffilmark{8},
D.~Hanna\altaffilmark{2},
J.~Holder\altaffilmark{19},
T.~B.~Humensky\altaffilmark{20},
C.~A.~Johnson\altaffilmark{21},
P.~Kaaret\altaffilmark{22},
P.~Kar\altaffilmark{1},
M.~Kertzman\altaffilmark{23},
D.~Kieda\altaffilmark{1},
M.~Krause\altaffilmark{9},
F.~Krennrich\altaffilmark{13},
S.~Kumar\altaffilmark{19},
M.~J.~Lang\altaffilmark{18},
G.~Maier\altaffilmark{9},
S.~McArthur\altaffilmark{11},
A.~McCann\altaffilmark{2},
K.~Meagher\altaffilmark{24},
P.~Moriarty\altaffilmark{18},
R.~Mukherjee\altaffilmark{14},
T.~Nguyen\altaffilmark{24},
D.~Nieto\altaffilmark{20},
R.~A.~Ong\altaffilmark{6},
A.~N.~Otte\altaffilmark{24},
N.~Park\altaffilmark{25},
V.~Pelassa\altaffilmark{4},
M.~Pohl\altaffilmark{8,9},
A.~Popkow\altaffilmark{6},
E.~Pueschel\altaffilmark{5},
J.~Quinn\altaffilmark{5},
K.~Ragan\altaffilmark{2},
P.~T.~Reynolds\altaffilmark{26},
G.~T.~Richards\altaffilmark{24},
E.~Roache\altaffilmark{4},
C.~Rulten\altaffilmark{16},
M.~Santander\altaffilmark{14},
G.~H.~Sembroski\altaffilmark{11},
K.~Shahinyan\altaffilmark{16},
D.~Staszak\altaffilmark{25},
I.~Telezhinsky\altaffilmark{8,9},
J.~V.~Tucci\altaffilmark{11},
J.~Tyler\altaffilmark{2},
S.~P.~Wakely\altaffilmark{25},
O.~M.~Weiner\altaffilmark{20},
A.~Weinstein\altaffilmark{13},
A.~Wilhelm\altaffilmark{8,9} \&
D.~A.~Williams\altaffilmark{21} (VERITAS Collaboration), 
S.~Fegan\altaffilmark{28}, 
B. Giebels\altaffilmark{28} \&
D. Horan\altaffilmark{28} (\emph{Fermi}-LAT Collaboration), 
A. Berdyugin\altaffilmark{27}, 
J. Kuan\altaffilmark{20}, 
E. Lindfors\altaffilmark{27}, 
K. Nilsson\altaffilmark{29}, 
A. Oksanen\altaffilmark{30}, 
H. Prokoph\altaffilmark{31}, 
R. Reinthal\altaffilmark{27}, 
L. Takalo\altaffilmark{27} \&  
F. Zefi\altaffilmark{28}
}

\altaffiltext{1}{Department of Physics and Astronomy, University of Utah, Salt Lake City, UT 84112, USA}
\altaffiltext{2}{Physics Department, McGill University, Montreal, QC H3A 2T8, Canada}
\altaffiltext{3}{Department of Physics, Washington University, St. Louis, MO 63130, USA, \mbox{errando@physics.wustl.edu}}
\altaffiltext{4}{Fred Lawrence Whipple Observatory, Harvard-Smithsonian Center for Astrophysics, Amado, AZ 85645, USA}
\altaffiltext{5}{School of Physics, University College Dublin, Belfield, Dublin 4, Ireland}
\altaffiltext{6}{Department of Physics and Astronomy, University of California, Los Angeles, CA 90095, USA}
\altaffiltext{7}{Argonne National Laboratory, 9700 S. Cass Avenue, Argonne, IL 60439, USA}
\altaffiltext{8}{Institute of Physics and Astronomy, University of Potsdam, 14476 Potsdam-Golm, Germany}
\altaffiltext{9}{DESY, Platanenallee 6, 15738 Zeuthen, Germany}
\altaffiltext{10}{Astronomy Department, Adler Planetarium and Astronomy Museum, Chicago, IL 60605, USA}
\altaffiltext{11}{Department of Physics and Astronomy, Purdue University, West Lafayette, IN 47907, USA}
\altaffiltext{12}{Department of Physics and Center for Astrophysics, Tsinghua University, Beijing 100084, China.}
\altaffiltext{13}{Department of Physics and Astronomy, Iowa State University, Ames, IA 50011, USA}
\altaffiltext{14}{Department of Physics and Astronomy, Barnard College, Columbia University, NY 10027, USA,  \mbox{muk@astro.columbia.edu}}
\altaffiltext{15}{Department of Astronomy and Astrophysics, 525 Davey Lab, Pennsylvania State University, University Park, PA 16802, USA}
\altaffiltext{16}{School of Physics and Astronomy, University of Minnesota, Minneapolis, MN 55455, USA}
\altaffiltext{17}{Department of Physics, California State University - East Bay, Hayward, CA 94542, USA}
\altaffiltext{18}{School of Physics, National University of Ireland Galway, University Road, Galway, Ireland}
\altaffiltext{19}{Department of Physics and Astronomy and the Bartol Research Institute, University of Delaware, Newark, DE 19716, USA}
\altaffiltext{20}{Physics Department, Columbia University, New York, NY 10027, USA}
\altaffiltext{21}{Santa Cruz Institute for Particle Physics and Department of Physics, University of California, Santa Cruz, CA 95064, USA}
\altaffiltext{22}{Department of Physics and Astronomy, University of Iowa, Van Allen Hall, Iowa City, IA 52242, USA}
\altaffiltext{23}{Department of Physics and Astronomy, DePauw University, Greencastle, IN 46135-0037, USA}
\altaffiltext{24}{School of Physics and Center for Relativistic Astrophysics, Georgia Institute of Technology, 837 State Street NW, Atlanta, GA 30332-0430}
\altaffiltext{25}{Enrico Fermi Institute, University of Chicago, Chicago, IL 60637, USA}
\altaffiltext{26}{Department of Physical Sciences, Cork Institute of Technology, Bishopstown, Cork, Ireland}
\altaffiltext{27}{Tuorla Observatory, Department of Physics and Astronomy, University of Turku, Finland}
\altaffiltext{28}{Laboratoire Leprince-Ringuet, Ecole polytechnique, CNRS/IN2P3, Universit\'{e} Paris-Saclay, 91128, Palaiseau, France, \mbox{sfegan@llr.in2p3.fr}, \mbox{zefi@llr.in2p3.fr}}
\altaffiltext{29}{Finnish Centre for Astronomy with ESO, University of Turku, Finland}
\altaffiltext{30}{Nyrola observatory, Jyvaskylan Sirius ry, Finland}
\altaffiltext{31}{Department of Physics and Electrical Engineering, Linnaeus University, 351 95 V\"{a}xj\"{o}, Sweden}

\begin{abstract}
\b2 is a BL~Lac-type blazar that was first detected at TeV energies by the MAGIC atmospheric Cherenkov telescopes, and subsequently confirmed by the VERITAS observatory with data collected between 2009 and 2012. 
In 2014 February 08, VERITAS detected a large-amplitude flare from \b2 during routine monitoring observations of the blazar 1ES~1218+304, located in the same field of view. The TeV flux reached 2.4 times the Crab Nebula flux with a variability timescale of $<3.6$\,h. 
Multiwavelength observations with {\it Fermi}-LAT, {\it Swift}, and the Tuorla observatory revealed a correlated high GeV flux state and no significant optical counterpart to the flare, with a spectral energy distribution where the gamma-ray luminosity exceeds the synchrotron luminosity. When interpreted in the framework of a one-zone leptonic model, the observed emission implies a high degree of beaming, with Doppler factor $\delta > 10$, and an electron population with spectral index $p<2.3$.

\end{abstract}


\section{Introduction}
Extreme flux variability is one of the defining properties of the blazar class of active galactic nuclei, appearing at all wavelengths over a wide range of timescales. Flares with amplitudes up to hundred times the quiescent flux and variability timescales as short as 3 minutes have been observed at TeV energies \citep[$E\gtrsim 0.1\,\mathrm{TeV}$; see, e.g.,][]{hess_2155}. 
To date, six flaring BL~Lac-type blazars have been detected to exceed the flux of the Crab Nebula 
\citep[$1\, \mathrm{Crab}=\left(2.1 \pm 0.2\right)\times 10^{-10}$\,cm$^{-2}$\,s$^{-1}$ at $E>0.2\,\mathrm{TeV}$,][]{hillas85} at TeV energies. 
The large signal statistics obtained during bright flares enable flux-variability studies on minute timescales, resulting in tighter constraints on the size and location of the gamma-ray emitting region \citep[see, e.g.,][]{begelman} and probing the particle acceleration and cooling mechanisms in blazar jets \citep[see, e.g.,][]{bykov}.

\begin{table*}[htpb]
\center
\caption{Summary of the VERITAS and \lat\ results from observations of \b2 in different epochs. The VERITAS upper limit is computed at 95\% c.l. assuming a power-law spectrum with index $\Gamma = 3.0$.}
\begin{tabular}{lclccc}
\hline\hline
Instrument  & Energy range    & Dates & Live time & Significance & Flux [cm$^{-2}$\,s$^{-1}$] \\
\hline
VERITAS     & $>0.2$\,TeV    & {2013 Jan 06} -- {2013 May 12} (MJD 56298--56424)& 631\,min & $8.8\sigma$ & $\left(6.0\pm1.2\right)\times 10^{-12}$ \\
          &     & {2013 Feb 07} (MJD 56330) & 25\,min & $10.5\sigma$ & $\left(5.1\pm1.0\right)\times 10^{-11}$ \\
     &      & {2014 Jan 29} -- {2014 May 25} (MJD 56686--56802) & 748\,min & $23.6\sigma$ & $\left(2.4\pm0.2\right)\times 10^{-11}$ \\
          &    & {2014 Feb 08} (MJD 56696) & 45\,min &$46.5\sigma$ & $\left(5.0\pm0.1\right)\times 10^{-10}$ \\
          &    & {2014 Feb 09} (MJD 56697) & 25\,min &$1.6\sigma$ & $ < 1.4  \times 10^{-11}$ \\
          \hline
\lat & 0.1--500\,GeV  & {2013 Jan 06} -- {2013 May 12} (MJD 56298--56424) &  & $28.8\sigma$ &  $\left(6.8\pm 0.7\right)\times 10^{-8}$\\         
 &  & {2014 Jan 01} -- {2014 May 25} (MJD 56658--56802) & & $34.5\sigma$ &  $\left(1.0\pm0.1\right)\times 10^{-7}$\\         
 &  & {2014 Feb 05} -- {2014 Feb 09} (MJD 56693--56696) & & $17.4\sigma$ &  $\left(4.4\pm0.7\right)\times 10^{-7}$\\         
\hline\hline
\label{obs}
\end{tabular}
\end{table*}
This paper describes a large-amplitude gamma-ray flare from the blazar \b2 detected on UT date 2014 Feb 08, and compares its broadband properties to long-term observations of the source with VERITAS (TeV energies), \lat\ (GeV energies; $0.1\lesssim E \lesssim 100\,\mathrm{GeV}$), and the Tuorla optical observatory.
\b2 
(R.A. = $12^{\rm h}17^{\rm m}52^{\rm s}$,
decl. = $+30^{\circ}07^{\prime}00\!^{\prime\prime}1$, J2000),
also known as ON~325 or 1ES~1215+303, was first detected at TeV energies by MAGIC~\citep{magic_1215}. At GeV energies it is associated with 3FGL~J1217.8+3007 \citep{3fgl}. 
There is some uncertainty in the distance to this source, with values of $z = 0.130$ \citep{akiyama03} and $z=0.237$ \citep{lanzetta93} being quoted for its spectroscopic redshift. Based on the location of its synchrotron peak, \b2 has been either classified as an intermediate \citep[IBL,][]{nieppola06} or high-frequency peaked BL~Lac \citep[HBL,][]{3lac}.

Throughout this paper we assume a Friedmann universe with $H_0=67.7$\,km\,s$^{-1}$\,Mpc$^{-1}$, $\Omega_{m} = 0.309$ and $\Omega_{\lambda} = 0.691$. All distance-dependent quantities are calculated assuming a redshift  $z=0.130$ ($d_L=630$\,Mpc) for \b2.
Measurement uncertainties are statistical only unless indicated otherwise.

\section{VERITAS Observations} 
\label{s:vts}
VERITAS (Very Energetic Radiation Imaging Telescope Array System) is 
an array of four imaging atmospheric Cherenkov telescopes located at the Fred Lawrence Whipple Observatory in southern Arizona, USA. VERITAS operates by recording Cherenkov light from particle showers initiated by gamma rays in the upper atmosphere and is sensitive to gamma-ray energies from about 85\,GeV to more than 30\,TeV \citep{holder_veritas}.

Table~\ref{obs} summarizes the VERITAS observations and results on \b2.
Observations were made in ``wobble'' pointing mode \citep{fomin94} considering the presence of another TeV source in the field of view (1ES~1218+304, offset $0.76^\circ$ from \b2)  as described in \citet{veritas_1215}.
Data were processed using standard VERITAS analysis pipelines 
 \citep[][]{ver_analysis2,ver_analysis}.
The energy threshold of the analysis is 200\,GeV, 
with a systematic uncertainty of 20\% on the energy estimation.

\begin{figure*}[htbp]
\centering
\plottwo{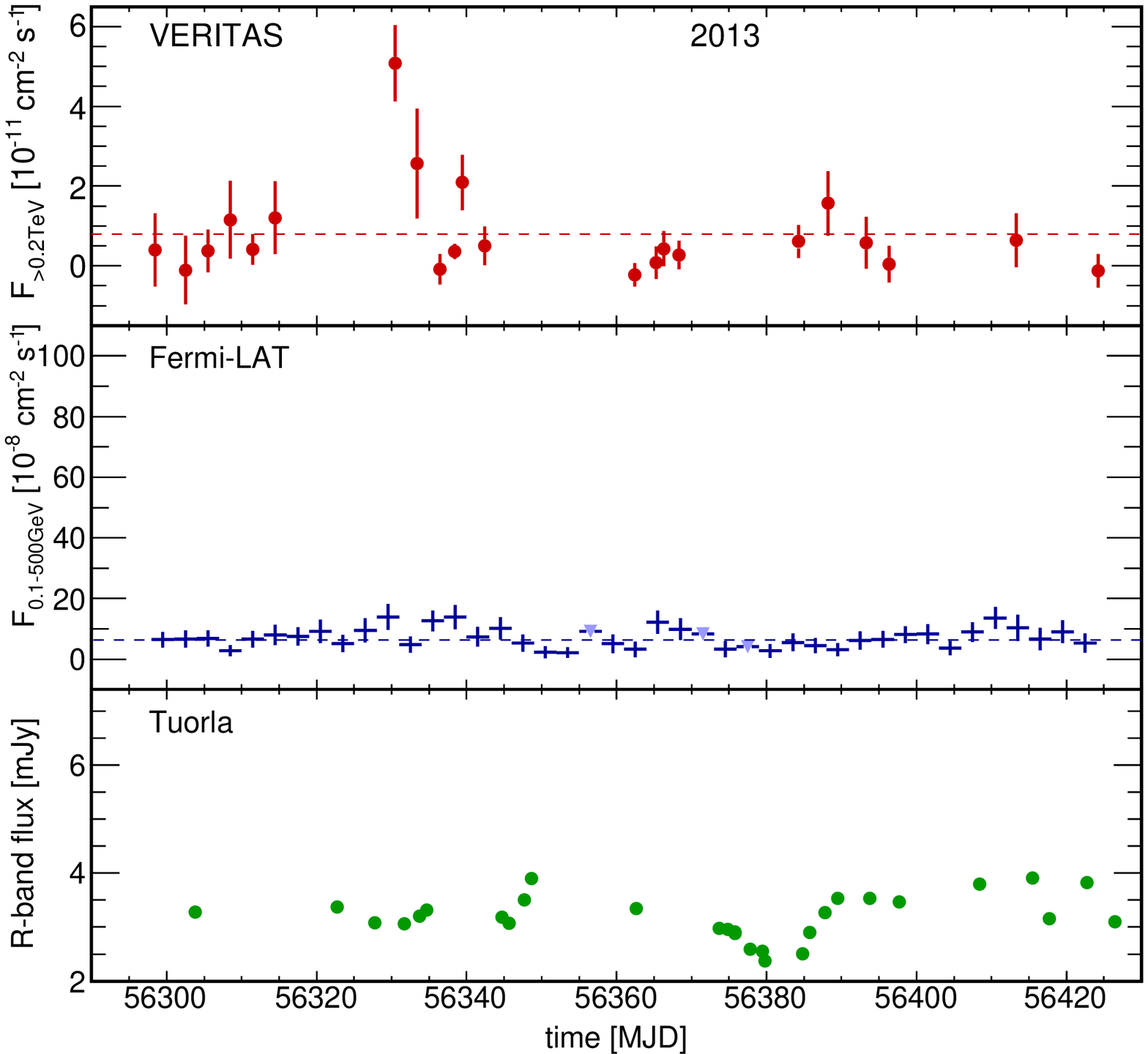}{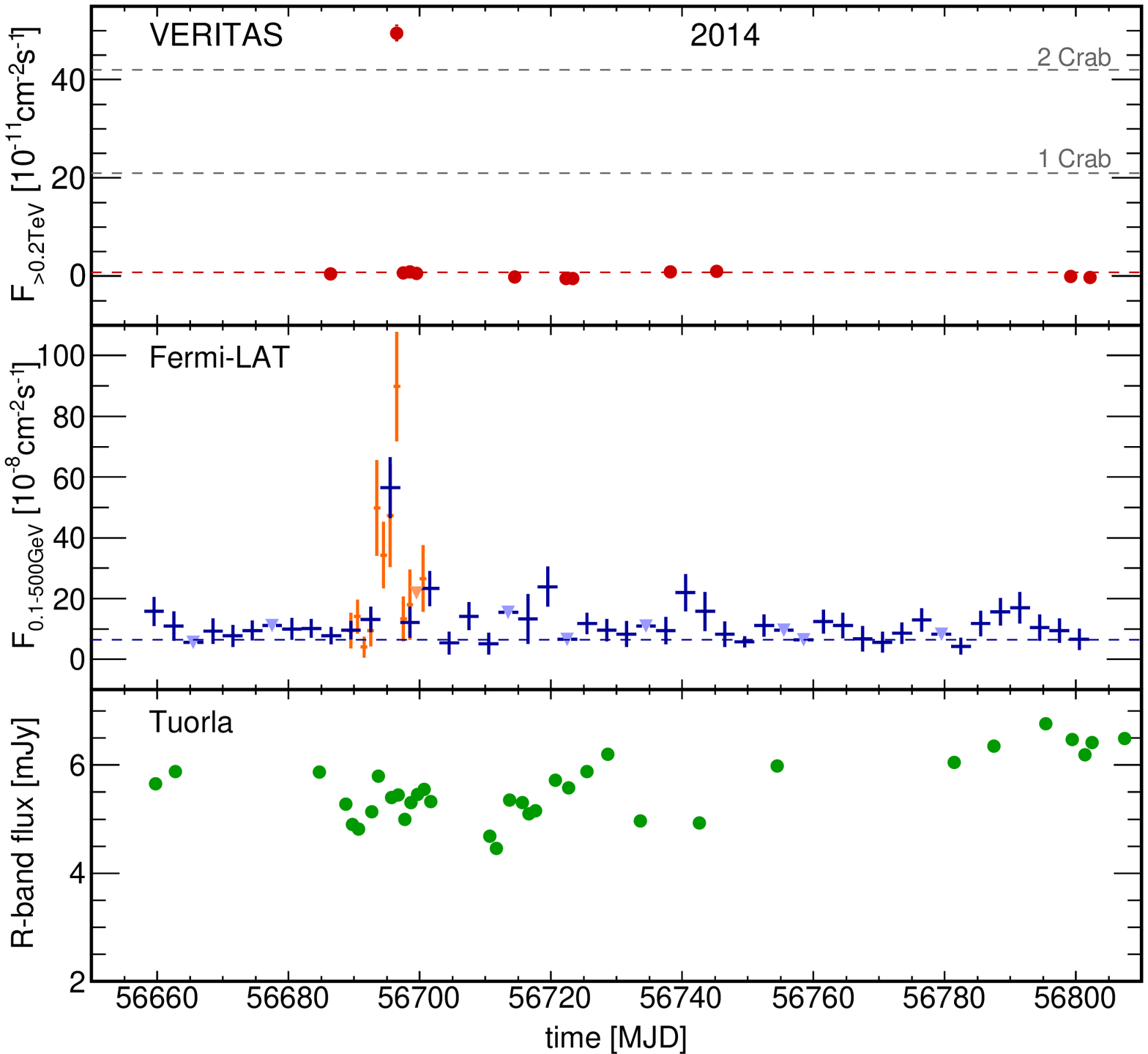}
\caption{TeV (\emph{top}), GeV (\emph{middle}), and optical (\emph{bottom}) light curves of \b2 in 2013 (\emph{left panel}) and 2014 (\emph{right panel}). Fluxes are calculated in 1-day bins for VERITAS. \lat\ fluxes are calculated with 3-day integration bins (blue crosses) and 1-day bins (orange crosses) around the time of the 2014 flare. Down-pointing triangles indicate 95\% c.l. upper limits derived from the \lat\ data for time bins with signal  smaller than 2$\sigma$. The yearly-averaged TeV flux in 2011 \citep[$8.0 \times 10^{-12}\,\mathrm{cm}^{-2}\,\mathrm{s}^{-1}$, ][]{veritas_1215} is shown by a red-dashed line, and a blue-dashed line indicates the average GeV flux from \citet{3fgl}. Statistical errors on the Tuorla optical fluxes are smaller than the data points.
}
\label{lc2013}
\label{lc2014}
\end{figure*}

A TeV flare from \b2 was detected 
in 2013 Feb 07 (MJD 56330, Figure~\ref{lc2013}) with flux $F_{>0.2\,\mathrm{TeV}}=\left(5.1 \pm 1.0_\mathrm{stat} \pm 1.0_\mathrm{sys} \right) \times 10^{-11}\,\mathrm{cm}^{-2}\,\mathrm{s}^{-1}$, or 0.24\,Crab. 
The measured gamma-ray spectrum is compatible with a power-law $\left(\mathrm{d}N/\mathrm{d}E=N_0 \cdot E^{-\Gamma}\right)$ with photon index $\Gamma = 3.7 \pm 0.7_\mathrm{stat} \pm 0.4_\mathrm{sys}$, in line with $\Gamma = 3.6 \pm 0.4$ reported in \citet{veritas_1215} and $\Gamma = 3.0 \pm 0.1$ from \citet{magic_1215}. A fit of the decaying phase of the flare (MJD 56330-56639) to a function $F\left(t\right) = F_0 \left(1+2^{-(t-t_0)/t_\mathrm{var}}\right)$ results in an upper limit on the flux halving time of $t_\mathrm{var} < 52$\,h at a 90\% confidence level (c.l.). 

A brighter subsequent flare from \b2 was observed 
on 2014 Feb 08 (MJD 56696, Figure~\ref{lc2014}) with  flux $F_{>0.2\,\mathrm{TeV}}= ( 5.0 \pm 0.1_{\mathrm{stat}} {}^{+4.0_\mathrm{sys}} _{-1.0_\mathrm{sys}} ) \times 10^{-10}\,\mathrm{cm}^{-2}\,\mathrm{s}^{-1}$, 
or 2.4\,Crab. The reconstructed energy spectrum is compatible with a power-law with photon index $\Gamma = 3.1 \pm 0.1_\mathrm{stat} \pm 0.6_\mathrm{sys}$ between 0.2 and 2\,TeV (Figure~\ref{sed}). The observations targeted 1ES~1218+304 and had a mean zenith distance of $27^{\circ}$, accumulating 45\,min of live-time exposure.
On that night, a high-cloud layer  at an altitude of $11.2$\,km a.s.l. was measured by an onsite Vaisala CL51 ceilometer. 
On average, 30\% of the Cherenkov light output in particle showers initiated by 200\,GeV gamma rays is produced above 11.2\,km \citep[see, e.g.,][]{1941RvMP...13..240R}. This fraction decreases with increasing gamma-ray energy \citep[see, e.g.,][]{2003WSAA...11.....W}.
If all Cherenkov light emitted above the cloud layer is lost, VERITAS would underestimate the energy of incoming gamma rays by $\sim 30\%$, which added to the 20\% systematic uncertainty on the energy estimation results in the increased systematic error on the gamma-ray flux and spectral index measured in 2014 Feb 08.
The large signal statistics during the flare allow flux measurements in 5-minute time bins (Figure~\ref{5min}). No significant flux variability was detected during the 45\,min exposure, with the light curve deviating from a constant flux hypothesis at a level of 2.8 standard deviations. Observations on the next night (2014 Feb 09) did not show an elevated flux from \b2  (Table~\ref{obs}), implying a 90\% c.l. limit on the flux halving time of 
$t_\mathrm{var} < 3.6$\,h.

\section{Fermi-LAT Observations}
\label{s:lat}
The Large Area Telescope (LAT) is a pair-conversion gamma-ray telescope on board the {\it{Fermi}} satellite covering energies from about 20\,MeV to more than 500\,GeV \citep{Atwood}. 
Table~\ref{obs} summarizes the \lat\ observations and results on \b2. 
Data were analyzed using the unbinned likelihood analysis in LAT ScienceTools ({\tt v10r0p5})  with {\tt P8R2\_SOURCE\_V6} instrument response functions, selecting photons with energy $100\,\mathrm{MeV}<E<500\,\mathrm{GeV}$ in a circular region of $10^{\circ}$ radius centered on the position of \b2. The energy spectrum of \b2 was modeled with a power law. Further analysis details and standard quality cuts followed \citet{3fgl}.
Light curves were derived by dividing the data in bins of one and three days duration. 

A clear flux peak is seen coinciding with the VERITAS-detected flare of 2014 Feb 08 (Figure~\ref{lc2013}), followed by a rapid decay that constrains the flux halving time to $t_{\mathrm{var}} < 8.9$\,h at 90\% c.l. (Figure~\ref{5min}). The GeV spectrum shows some evidence of hardening ($2.2 \sigma$), going from an averaged $\Gamma_\mathrm{GeV} = 1.92 \pm 0.04$ during the 2014 campaign to $\Gamma_\mathrm{GeV} = 1.70 \pm 0.09$ in the four days of highest GeV flux (MJD 56693-56696).
In 2013,  the LAT light curve shows no significant flux variability (Figure~\ref{lc2013}). However, the same TeV to GeV flare amplitude ratio seen in 2014 can be accomodated within the error bars of the 2013 LAT light curve.

\section{Swift observations}
\label{sec:xray}
An observation by the {\it Swift}
Observatory (ObsId 00031906012) was carried out one day after the VERITAS-detected flare (Figure~\ref{5min})  
with an exposure of 1.97\,ks. X-ray Telescope  \citep[XRT, $0.2-10$\,keV,][]{2005SSRv..120..165B} data  were obtained in photon-counting 
mode and processed with the
 {\tt xrtpipeline} tool (HEASOFT 6.16).
The exposure shows a stable 
source-count rate of 
$\sim 0.3\,{\rm s^{-1}}$, suggesting negligible pile-up effects.

The spectrum was rebinned to
have at least 20 counts per bin,
ignoring channels with energy below 0.3\,keV, and fit using PyXspec v1.0.4
\citep{1996ASPC..101...17A}.
An absorbed power law with column density $N_{\rm H}=1.68\times10^{20}\,{\rm cm^{-2}}$ \citep{2005A&A...440..775K} and photon index $\Gamma_{\rm X}=2.54\pm0.07$ gives a good description of the spectral data ($P(\chi^2)=0.42$). 
The unabsorbed flux is $F_{0.3-10\,{\rm keV}}=\left(1.28 \pm 0.05 \right)\times 10^{-11}\,{\rm erg}\,{\rm cm^{-2}}\,{\rm s^{-1}}$.

To analyze the {\it Swift}-UVOT data ($E\sim6.0$\,eV), source counts were extracted from an aperture
of 5.0 arcsec radius around the source. Background counts were
taken from four neighboring regions with equal radius.
Magnitudes were  computed using the
{\tt{uvotsource}} tool (HEASOFT v6.16), 
corrected for extinction according to \citet{2009ApJ...690..163R} 
using $E(B-V)$ from \citet{2011ApJ...737..103S}, and converted to fluxes following 
\citet{2008MNRAS.383..627P}.

\section{Optical observations}
Optical R-band data were obtained as part of the Tuorla blazar monitoring program \citep[http://users.utu.fi/kani/1m, ][]{takalo}. Observations were taken using a 35\,cm Celestron telescope attached to the KVA 60\,cm telescope (La Palma, Canary Islands, Spain) and the 50\,cm Searchlight Observatory Network telescope (San Pedro de Atacama, Chile). Data were analyzed using a semi-automatic pipeline developed at the Tuorla Observatory. 
The host galaxy flux of 1.0 mJy \citep{nilsson} was subtracted from the observed fluxes, and a correction for Galactic extinction was applied using values from \citet{2011ApJ...737..103S}.
The yearly-averaged optical flux of $\left(3.27 \pm 0.01\right)$\,mJy in year 2013 is similar to historical values dating back to 2003.\footnote{http://users.utu.fi/kani/1m/ON\_325\_jy.html} In 2014, \b2 appeared to be in a long-lasting high optical state, with average flux of $\left(5.56 \pm 0.02\right)$\,mJy.
No significant enhancement of the optical emission was detected in coincidence with the two gamma-ray flares reported in Sections \ref{s:vts} and \ref{s:lat}. 

\section{Discussion} 
\label{discussion}
With the data presented here and in \citet{veritas_1215}, VERITAS has published TeV observations of \b2 spanning over 50 nights between 2008 and 2014, finding no significant deviations from yearly-averaged fluxes other than the flares on 2013 Feb 07 and 2014 Feb 08 reported in this paper. 
These two TeV flares had amplitudes of $\sim 6$ and $\sim 60$ times the average quiescent flux from \b2, with associated
flux halving times of $\sim 52$ and $\sim 3.6$ hours, respectively.
Such large-amplitude, short-lived, isolated flares are not common in TeV-emitting blazars. Fast variability is typically measured during longer high-flux states in HBLs \citep[see, e.g.,][]{1959,501}, 
while some quasars and IBLs show short periods of TeV emission in epochs where multiple GeV flares are seen \citep{magic_1222,veritas-BLLac}.

In the following we summarize the main observational properties of the brightest flare of 2014 Feb 08 and interpret them in the framework of an homogeneous one-zone leptonic emission scenario:

(i) The measured flux above 0.2\,TeV was $\left(5.0 \pm 0.1 \right) \times 10^{-10}\,\mathrm{cm}^{-2}\,\mathrm{s}^{-1}$. 
This corresponds to an isotropic luminosity $L_\gamma = 1.7 \times 10^{46}$ erg~s$^{-1}$.  
To date, only four other blazars 
have episodically been observed to emit TeV radiation with luminosity exceeding $10^{46}$\,erg\,s$^{-1}$.
For comparison, the historical TeV blazar Mrk~421 
would have to exhibit a 35\,Crab flare to reach the luminosity of the \b2 outburst reported here.

(ii) 
A non-detection by VERITAS 24\,h after the flare 
indicates a flux halving time $t_\mathrm{var}<3.6$\,h at TeV energies. 
Causality implies that the observed variability timescale is related to the size ($R$) and Doppler factor ($\delta$) of the gamma-ray emitting region by
\begin{linenomath}
\begin{equation}
R \delta^{-1} \leq c\,t_{\mathrm{var}}/(1+z) = 3.4 \times 10^{14}\,\mathrm{cm},
\label{causality}
\end{equation}
\end{linenomath}

(iii) The TeV flare was accompanied by a significant GeV flare measured by \lat\ that extended over four days and displayed some evidence for spectral hardening, with $\Gamma_\mathrm{GeV} = 1.70 \pm 0.09$. 

(iv) Optical observations did not show enhanced emission in coincidence with the GeV and TeV flare, although the overall optical flux in 2014 was approximately two times brighter than in previous years.

(v) Non-detections by {\it Swift}-BAT\footnote{ \url{http://swift.gsfc.nasa.gov/results/transients/weak/QSOB1215p303/}} (15-50\,keV) and MAXI\footnote{ \url{http://maxi.riken.jp/mxondem/}} (4-10\,keV) on the day of the TeV flare (MJD 56696) can be interpreted as a limit on the hard X-ray flux of the order of $\nu_\mathrm{x} F_{\nu_\mathrm{x}} \lesssim 2 \times 10^{-10}$\,erg\,cm$^{-2}$\,s$^{-1}$ \citep{bat,maxi}. This effectively limits the peak synchrotron luminosity to 
\begin{linenomath}
\begin{equation}
L_\mathrm{syn} \leq 10^{46}\,\mathrm{erg}~\mathrm{s}^{-1}.
\label{Lsy}
\end{equation}
\end{linenomath}

(vi) No change in the 15\,GHz radio brightness of \b2 was seen in the OVRO light curves in coincidence or after the TeV flare.\footnote{ \url{http://www.astro.caltech.edu/ovroblazars/data/data.php?page=data\_return\&source=J1217+3007}} \b2 is in fact in the lower third of the OVRO sample in terms of radio flux variability \citep{2014MNRAS.438.3058R}. 

(vii) {\it Swift}-XRT data taken 24\,h after the flare showed an X-ray flux comparable with historical average values \citep{magic_1215,veritas_1215}, although the TeV flux was back to a quiescent level at that point.

\begin{figure}
\centering
\includegraphics[width=1.0\columnwidth]{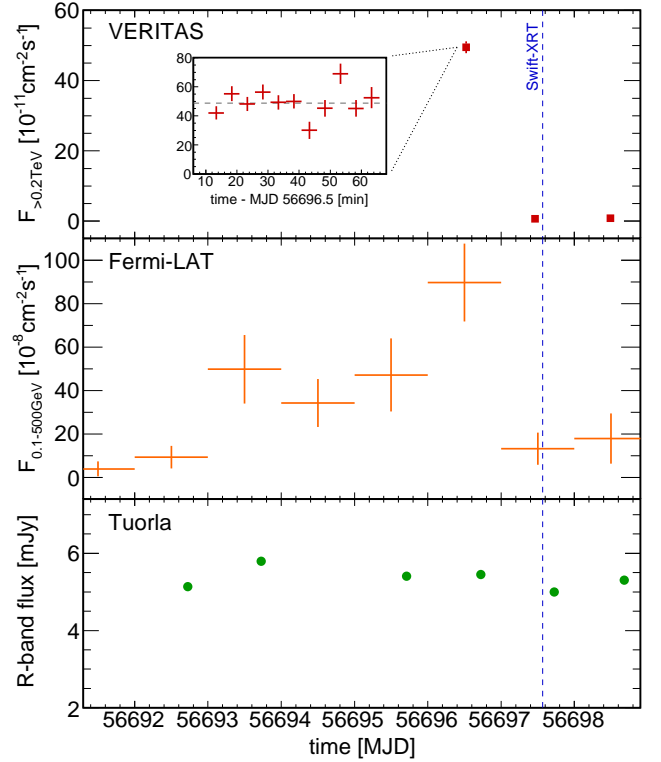}
\caption{ Same as Figure~\ref{lc2013} around the night of 2014 Feb 08 (MJD 56696). The top panel insert shows the TeV flux on MJD 56696 in 5-minute bins. A fit of the 5-minute binned TeV light curve to a constant flux (gray-dashed line) yields $P(\chi^2)=4.2 \times 10^{-3}$. A vertical blue-dashed line indicates the time of the {\it Swift}-XRT observation described in Section~\ref{sec:xray}.}
\label{5min}
\end{figure}

A lower limit on $\delta$ can be derived by estimating the required Doppler boosting for gamma rays with energy $E_\gamma$ to escape pair production on a co-spatial synchrotron photon field with density $F\left(E_0\right)$, where $E_0 = \left(m_e c^2\right)^2 \left(1+z \right)^{-2} \delta^2  E_\gamma^{-1}$.
For photons with $E_\gamma \sim 1$\,TeV measured by VERITAS the mean interaction energy for pair production is $E_0=76$\,eV.
Using the expression for optical depth from \citet{dondi95}, imposing $\tau_{\mathrm{\gamma\gamma}}\leq 1$, and estimating $F\left(E_0\right)$ from the {\it Swift}-XRT and UVOT measurements described in Section~\ref{sec:xray}
results in
\begin{linenomath}
\begin{eqnarray}
\delta & \geq & \left[ \frac{\sigma_\mathrm{T}d^2_\mathrm{L}\left(1+z\right)^{2\alpha}}{5hc^2} \frac{F\left(E_0\right)}{t_\mathrm{var}}\right]^{1/\left(4+2\alpha\right)}\,, \nonumber \\
\delta & \geq & 10.0\, ,
\label{delta}
\end{eqnarray}
\end{linenomath}
where $\sigma_\mathrm{T}$ is the Thomson cross section and $\alpha$ is the spectral index of the synchrotron emission around $E_0$. 
We note that the {\it Swift} observations 
were made 24\,h after the TeV flare (Figure~\ref{5min}). The lower limit on $\delta$ is still valid, however, as long as the density of synchrotron photons was not lower during the flare than that measured on the subsequent day.

\begin{figure}
\centering
\includegraphics[width=1.0\columnwidth]{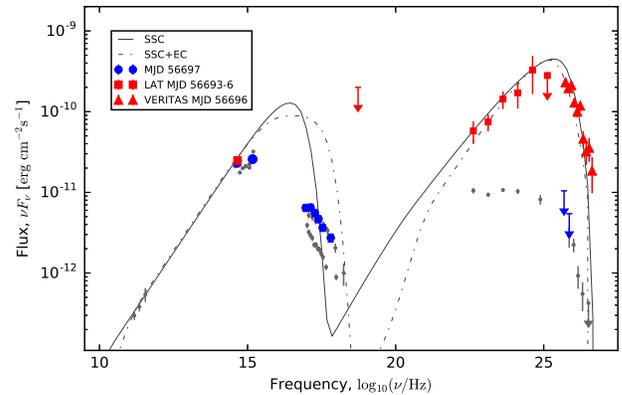}
\caption{Broadband SED of \b2 at different epochs. Red markers show the state of the source during the 2014 Feb 08 flare, including VERITAS (MJD 56696.52), \lat\ (MJD 56693-56696), {\it Swift}-BAT (MJD 56696), and Tuorla (MJD 56696.72) data. 
Blue markers show {\it Swift}-XRT and UVOT fluxes and VERITAS 95\% c.l. upper limits taken 24\,h after the flare. Gray markers show archival observations from \citet{veritas_1215}. The numerical SSC and SSC+EC models described in Section~\ref{discussion} are shown with a solid and a dashed gray line, respectively. Gamma-ray absorption by the extragalactic background light is applied to the models following \citet{ebl}.}
\label{sed}
\end{figure} 

The spectral energy distribution (SED) of \b2 during the flare is shown in Figure~\ref{sed}.  TeV emission can be explained by a fresh injection of relativistic electrons, where the injected perturbation propagates down in energy as the plasma cools, explaining the smaller amplitude of the GeV flare and the lack of optical variability \citep[see, e.g., ][]{2007A&A...462...29G}. 
Taking the radio spectrum from \citet{2004MNRAS.352..673A} and the R-band flux from the Tuorla  observatory we derive a radio-to-optical spectral index $\alpha_\mathrm{ro} =0.45$. 
If the cooling break\footnote{corresponding to emitting electron energies at which the radiative cooling and escape timescales are equal.} in the synchrotron SED happens beyond optical frequencies, as assumed in \citet{magic_1215} and \citet{veritas_1215} and typically observed in BL~Lac objects \citep{2010MNRAS.401.1570T}, $\alpha_\mathrm{ro}$ determines the power-law spectral index ($p$) of the emitting electrons \citep[see, e.g.,][]{1979rpa..book.....R}:
\begin{linenomath}
\begin{equation}
p=1+2\alpha_\mathrm{ro}\approx 1.9.
\label{p}
\end{equation}
\end{linenomath}
Beyond the cooling break, the electron distribution has to extend to Lorentz factors ($\gamma$) of the order
\begin{linenomath}
\begin{equation}
\gamma_\mathrm{max} \approx \left( 1+z\right) \delta^{-1} \dfrac{1\,\mathrm{TeV}}{m_e c^2} \geqslant 2.2 \times 10^5 \left(\delta/10\right)^{-1}
\label{gmax}
\end{equation}
\end{linenomath}
to produce the $\sim 1$\,TeV photons detected by VERITAS. In the simplest leptonic emission scenario, the high-energy component of the SED is produced via the synchrotron self-Compton mechanism \citep[SSC;][]{1992ApJ...397L...5M}. In an SSC scenario, the ratio between the synchrotron and inverse-Compton luminosities can be used to estimate the magnetic field. Following \citet{1996A&AS..120C.503G} and using equations~(\ref{Lsy}) and (\ref{delta}) to constrain  $L_\mathrm{syn}$ and $\delta$, we derive
\begin{linenomath}
\begin{eqnarray}
B &\simeq& \left(1+z \right)\, \delta^{-3}\left(\dfrac{2L_\mathrm{syn}^2}{L_\gamma c^3 t_\mathrm{var}}\right)^{1/2}, \nonumber \\
&\leqslant& 1.8\,\mathrm{G}\,\left( L_\mathrm{syn}/10^{46}\,\mathrm{erg\,s^{-1}}\right) \left(\delta/10\right)^{-3}.
\label{B}
\end{eqnarray}
\end{linenomath}

The scarcity of multiwavelength coverage simultaneous with the TeV flare, specially of the synchrotron component, leaves numerical modeling of the SED underconstrained.  However, even if modeling solutions are not unique, they
can be used to understand the level of kinetic and magnetic jet power required under different scenarios. 
We test the feasibility of a SSC scenario by using the stationary leptonic model of \citet{2013ApJ...768...54B}, fixing the jet viewing angle to $\delta^{-1}$ for simplicity. 
Models\footnote{E.g., $L_e=1.05\times10^{45}\,\mathrm{erg}\,\mathrm{s^{-1}}$, $q_e=1.9$, $\delta = \gamma_\mathrm{min}=40$, $\gamma_\mathrm{max}=10^5$, $B=0.03\,\mathrm{G}$, $R=1.3\times10^{16}\,\mathrm{cm}$, $\eta_\mathrm{esc} =1$, see \citet{2013ApJ...768...54B} for parameter definitions not included in the text.} within the parameter constraints from equations~(\ref{causality}--\ref{B}) reproduce the measured gamma-ray luminosity without overproducing the optical flux measured by the Tuorla observatory, 
and keeping $L_\mathrm{syn} \lesssim L_\gamma$ as constrained by the {\it Swift}-BAT non-detection (Figure~\ref{sed}).
These solutions would indicate an emitting region where the kinetic power of relativistic electrons ($L_\mathrm{e}$) exceeds the power carried by the magnetic field ($L_\mathrm{B}$) by a factor of $\sim 1200$.
This is typically the case in SSC modelling of TeV blazars \citep[see, e.g., ][]{veritas_1215}.
Higher values of $\delta$ would imply even higher $L_\mathrm{e}/L_\mathrm{B}$ ratios. 
Given the observational uncertainty in the shape of the synchrotron emission, we also explore a wider range of electron spectral indices than indicated in equation~(\ref{p}), finding that $p < 2.3$ is required to reproduce the hard GeV spectrum measured by \lat. 

The lack of observable thermal emission from the accretion disk and associated emission lines in \b2 supports an SSC emission scenario. However, the observed Compton dominance ($L_\gamma/ L_\mathrm{syn} \gtrsim 1$) typically points to external Compton models \citep[EC;][]{1993ApJ...416..458D} to explain the high-energy emission. Assuming an EC scenario, 
constraints on $\delta$ and the distance of the energy dissipation region from the black hole ($r_\mathrm{diss}$) can be derived assuming reasonable limits on the jet collimation and  luminosity of upscattered synchrotron photons. Following \citet{nalewajko} results in:
\begin{linenomath}
\begin{eqnarray}
\delta \left(r_\mathrm{diss}\right) &<& \left[  \dfrac{\left(1+z\right)\,r_\mathrm{diss}}{c\,t_\mathrm{var}}  \right]^{1/2}, \label{gt}\\
\delta \left(r_\mathrm{diss}\right) &>& \left[ \dfrac{9}{2} \dfrac{L_\gamma}{\zeta\left(r_\mathrm{diss}\right)L_\mathrm{d}} \right]^{1/8} 
 \left[\dfrac{\left(1+z\right)r_\mathrm{diss}}{2\,c\,t_\mathrm{var}}\right]^{1/4}, \label{ssc}
\end{eqnarray}
\end{linenomath}
where the accretion disk luminosity ($L_\mathrm{d}$) is assumed to be $4 \times 10^{43}$\,erg~s$^{-1}$ \citep{2010MNRAS.402..497G} and $\zeta\left(r_\mathrm{diss}\right)$ describes the composition of the external radiation fields. Equations~(\ref{gt}) and (\ref{ssc}) constrain the $\left(\delta,\, r_\mathrm{diss}\right)$ parameter space  with a marginal solution at $\delta > 19$ and $r_\mathrm{diss} > 1.2 \times 10^{17}$\,cm that would place the emitting blob beyond the broad-line region. 
A numerical EC model\footnote{$L_e=5\times10^{43}\,\mathrm{erg}\,\mathrm{s^{-1}}$, $q_e=1.9$, $\delta = \gamma_\mathrm{min}=40$, $\gamma_\mathrm{max}=10^5$, $B=0.3\,\mathrm{G}$, $R=10^{16}\,\mathrm{cm}$, $u_\mathrm{ext}=2\times10^{-6}\,\mathrm{erg}\,\mathrm{cm^{-3}}$, $T_\mathrm{ext}=10^{3}\,\mathrm{K}$, $\eta_\mathrm{esc} =1$, see \citet{2013ApJ...768...54B} for parameter definitions not included in the text.} \citep{2013ApJ...768...54B} with an external photon field described as blackbody emission with $T_\mathrm{ext}=10^3$\,K typical of hot dust can reproduce the SED with  $L_\mathrm{e}/L_\mathrm{B} \sim 1 $ (Figure~\ref{sed}).

Particle acceleration in relativistic shocks or through magnetic reconnection can  explain the short flux-variability timescale observed in \b2 \citep{sironi09,2013MNRAS.431..355G}. The hard electron spectrum ($ p \lesssim 2.3$) derived from the multiwavelength SED is usually obtained in semi-analytical calculations of relativistic shock acceleration 
\citep{2001MNRAS.328..393A}, but more
recent fully kinetic particle-in-cell simulations derive significantly softer spectra \citep{sironi09}. 
Magnetic reconnection events can produce harder electron spectra than those originating from shock acceleration \citep[][]{sironi14}, easily reproducing $p\sim1.9$ derived from the synchrotron spectrum of \b2. Recently, \citet{sironi15} have suggested that magnetic reconnection is a more viable scenario for particle acceleration in relativistic jets,  disfavoring shock models for their inability to simultaneously dissipate energy and accelerate particles beyond thermal energies. Efficient magnetic reconnection requires an emitting region in rough equipartition between particles and magnetic field  $\left(L_\mathrm{e}/L_\mathrm{B} \lesssim 1 \right)$. The EC scenario presented above does fulfill this condition, while our attempts to describe the observed SED with SSC models persistently resulted in particle-dominated emitting regions where the magnetization of the plasma would be too low for efficient magnetic reconnection to take place.
 
VERITAS will continue to monitor \b2. Events like the extreme flare of 2014 Feb 08 should be within the sensitivity reach of HAWC \citep{2015arXiv150804479L}. 
Future observations will show how frequent these extreme gamma-ray flares are and whether or not they are present in the majority of TeV blazars.

\acknowledgements
The authors thank Markus B{\"o}ttcher for valuable discussions about leptonic emission models, and David Sanchez for providing useful comments on the draft. RM  acknowledges support from the Alliance Program at \'Ecole Polytechnique and Columbia University.
VERITAS research is supported by grants from the U.S. Department of Energy Office of Science, the U.S. National Science Foundation and the Smithsonian Institution, and by NSERC in Canada. We acknowledge the excellent work of the technical support staff at the Fred Lawrence Whipple Observatory and at the collaborating institutions in the construction and operation of the instrument. The VERITAS Collaboration is grateful to Trevor Weekes for his seminal contributions and leadership in the field of VHE gamma-ray astrophysics, which made this study possible.
The {\it Fermi}-LAT Collaboration acknowledges generous ongoing support from a number of agencies and institutes that have supported both the development and the operation of the LAT as well as scientific data analysis. These include the National Aeronautics and Space Administration and the Department of Energy in the United States, the Commissariat \`{a} l'Energie Atomique and the Centre National de la Recherche Scientifique/Institut National de Physique Nucl\'{e}aire et de Physique des Particules in France, the Agenzia Spaziale Italiana and the Istituto Nazionale di Fisica Nucleare in Italy, the Ministry of Education, Culture, Sports, Science and Technology (MEXT), the High Energy Accelerator Research Organization (KEK) and Japan Aerospace Exploration Agency (JAXA) in Japan, and the K.A. Wallenberg Foundation, the Swedish Research Council, and the Swedish National Space Board in Sweden. Additional support for science analysis during the operations phase is gratefully acknowledged from the Istituto Nazionale di Astrofisica in Italy and the Centre National d'\`{E}tudes Spatiales in France.

\end{document}